     \documentclass[12pt]{article}   
     \newlength{\dinwidth}                       
     \newlength{\dinmargin}                      
\def\Journal#1#2#3#4{{#1} {\bf #2}, #3 (#4)}
\def\CPC{\em Comp. Phys. Commun.}

\def\NPB{{\em Nucl. Phys.} B}

\def\PRD{{\em Phys. Rev.} D}
\def\ZPC{{\em Z. Phys.} C}
\def\lsim{\mathrel{\rlap{\lower4pt\hbox{\hskip1pt$\sim$}}
    \raise1pt\hbox{$<$}}}                
\def\gsim{\mathrel{\rlap{\lower4pt\hbox{\hskip1pt$\sim$}}
    \raise1pt\hbox{$>$}}}                
\def\HER{{\sc Herwig~5.9}}
\def\EEBAR{$e^+e^-$}
\def\dd{\mathrm{d}}
\begin{document}
\vspace*{10mm}
\begin{center}  \begin{Large} \begin{bf}
Cluster Hadronization in \HER\\
  \end{bf}  \end{Large}
  \vspace*{5mm}
  \begin{large}
Alexander Kup\v{c}o\\
  \end{large}
\end{center}
Institute of Physics, Na Slovance 2, 182 21 Prague 8, Czech Republic \\
\begin{quotation}
\noindent
{\bf Abstract:} The \HER\ cluster hadronization model is briefly discussed
here. It is shown that  the model has peculiar behaviour when new heavy
baryon resonances are included  in the \HER\  particle table. New
fragmentation model is proposed to cure  this problem and simple
tuning of \HER\  with this new model has been made using event shapes
variables and identified particle momentum spectra in   \EEBAR\
interactions at LEPI. Finally, the predictions of the two
hadronization  models are compared.
\end{quotation}
\section{Introduction}

Fragmentation of quarks and gluons into hadrons is a typical
non-perturbative phenomenon and phenomenological methods
must be used to model it. The three basic models are available
at present: independent fragmentation ({\sc
Isajet},~\cite{IsajetMan}), string fragmentation ({\sc
Jetset},~\cite{PythiaMan}), and cluster hadronization ({\sc
Herwig},~\cite{HerwigMan}).

The independent fragmentation model does not take into account the
colour connections and it is also less successful than other two
models. The concept of colour connection is essential for the string
and cluster hadronization models. {\sc Jetset} gives better agreement
with the experimental data then {\sc Herwig} but  also contains a
large number of parameters which can be tuned.  Nevertheless, {\sc
Herwig}, with its few-parametric cluster hadronization model,
represents main alternative to the string models.

In this paper it is shown that the behaviour of \HER\  in relation to
the light quark sector ($u$, $d$, $s$) does not correspond to the
naive view of cluster hadronization in that the introduction of the
new baryon cluster decay channels does not lead to an increase in the
predicted proton yield. In fact, the proton multiplicity decreases in
comparison to predictions made with the default particle list.

This article reviews the \HER\  hadronization model and pins down the
origin of the peculiar behaviour. A new hadronization model is
proposed to cure the problem. To make reliable comparison between the
two models, it was necessary to tune the new one. Finally, the
differences between the models are discussed. Wherever possible the
names of variables are taken from \HER\  notation. They are indicated
by \texttt{typewriter} font.

This work is important for analyses at the HERA experiments in that a
reliable theoretical description of the baryon sector needs to be
provided. Although there have so far been very few experimental
results on baryon yield from H1 and ZEUS, interesting observations are
already being made on strangeness suppression~\cite{zeusk0, h1k0} and
baryon number propagation~\cite{ppbar} and, with the high statistics
now available to the experiments, this will soon become an
experimentally productive area.
\section{Hadronization in \HER}

Colourless clusters are formed from colour connected
quarks,~\cite{March88}.  They consist of quark-antiquark (meson-like
clusters), quark-diquark (baryon-like), or antiquark-antidiquark
(anti\-baryon-like) pairs .  Only the meson-like clusters can occur in
\EEBAR\ interactions.  The basic idea of the model is that the
clusters decay according to the phase space available to the decay
products.
\subsection{\HER\  parameters and variables}

The most important \HER\  parameters and variables relevant to cluster
model (subroutine \texttt{HWCHAD}) are described in this
subsection. The values shown here are the default \HER\  settings.

The a priori weights for quarks $d$, $u$, $s$, $c$, $b$, $t$, and
diquarks (in that order) are assumed to be equal
\begin{equation}
  \label{PWT}
  \texttt{PWT(1)}=\cdots=\texttt{PWT(7)}=1\,.
\end{equation}
Weights for diquarks of different flavour compositions are also
calculated from these and the result is similarly stored in the array
\texttt{PWT}.

Parameters \texttt{CLMAX=3.35} and \texttt{CLPOW=2} determine the
maximum allowed mass of the cluster
\begin{equation}
  \label{MaxMass}
  M_{max}^{\tiny{\texttt{CLPOW}}}=\texttt{CLMAX}^{\tiny{\texttt{CLPOW}}}+
    \left(\texttt{RMASS(ID1)}+
    \texttt{RMASS(ID2)}\right)^{\tiny{\texttt{CLPOW}}}\,,
\end{equation}
where \texttt{RMASS(ID1)} and \texttt{RMASS(ID2)} are masses of the
quarks from which the cluster is formed. If the cluster is too heavy it is
first cut in two by creation of quark antiquark pair (subroutine
\texttt{HWCCUT}).

All particles are grouped according to their flavour content and their
internal codes  \texttt{IDHW} are stored in the one-dimensional array
\texttt{NCLDK}. The first (and also the lightest) hadron of a given
flavour (\texttt{ID1,ID2}) has position \texttt{LOCN(ID1,ID2)} in the
array \texttt{NCLDK}. The other particles with the same flavour are
stored from this position and their number is \texttt{RESN(ID1,ID2)}.
The masses of the lightest hadrons are put in \texttt{RMIN(ID1,ID2)}.
Another one-dimensional array \texttt{CLDKWT} stores spin weights and
mixing weights. For every flavour these weights are normalised by the
maximum weight.

Individual weights are stored for all particles in the table
\texttt{SWTEF}. This weights are set to one in the default \HER\
version.  There are also parameters that set individual weights for
group of particles in \HER, for example \texttt{DECWT} is the weight
for baryons in decuplet.
\subsection{Implementation of cluster hadronization model (\texttt{HWCHAD})}

If the mass \texttt{EM0} of cluster with flavour content
(\texttt{ID1,ID3}) is equal to the mass of the lightest
hadron with the same flavour then the cluster is identified with this
hadron. In the other cases the cluster decays into two hadrons.

To begin with, the channel with the lightest decay products is chosen. If
the sum of their masses (\texttt{EM1},~\texttt{EM2}) is bigger than
the cluster mass \texttt{EM0} then cluster decays by $\pi^0$ emission.
In the other cases the phase space \texttt{PCMAX} of this decay is
used as a maximum phase space weight
\begin{equation}
  \label{pcmax}
  \texttt{PCMAX}=\sqrt{\texttt{EM0}^2-(\texttt{EM1}+\texttt{EM2})^2}\cdot
                 \sqrt{\texttt{EM0}^2-(\texttt{EM1}-\texttt{EM2})^2}.
\end{equation}

The hadronization of the cluster proceeds in the following three steps.
\begin{enumerate}
\item The (di)quark anti(di)quark pair of the flavour \texttt{ID2} is
  created from vacuum according to the probability
  \begin{equation}
  \label{Pfl}
  P_{flavour}(\texttt{ID2}) = \texttt{PWT(ID2)}/
    \sum_{\texttt{I}}\texttt{PWT(I)},
  \end{equation}
where the sum is over all allowed flavours.
\item One hadron with flavour (\texttt{ID1},\texttt{ID2}) is randomly
  chosen from the iso-flavour table \texttt{NCLDK} and this particle is 
  accepted according to its weight \texttt{CLDKWT}.  If the particle
  is not accepted then another hadron with the same flavour content
  (\texttt{ID1},\texttt{ID2}) is taken from the table and so on. The
  corresponding probability that particle 
  $\texttt{IR1}=\texttt{NCLDK}(\texttt{I1})$ is chosen is then
  \begin{equation}
  \label{Ps}
    P_{spin}(\texttt{IR1}|\texttt{ID1}, \texttt{ID2}) =
    \frac{\texttt{CLDKWT(I1)}}
    {\sum_{\tiny{\texttt{LOCN(ID1,ID2)}}}^
    {\tiny{\texttt{LOCN+RESN-1}}}\texttt{CLDKWT(I)}}\ .
  \end{equation}
  The same procedure is applied to the hadrons with flavour
  (\texttt{ID2},\texttt{ID3}).
\item
  Once the decay channel has been chosen it is accepted according to 
  its phase space \texttt{PCM} (given by formula~(\ref{pcmax})) with 
  probability
  \begin{equation}
  \label{Pps}
    P_{phase}=
    \sqrt{\texttt{SWTEF(IR1)}}\cdot\sqrt{\texttt{SWTEF(IR2)}}\cdot
    \texttt{PCM}\,/\,\texttt{PCMAX}\ .
  \end{equation}
  If the decay mode is not accepted the algorithm starts again from
  the first point (i.e. from creation of quark antiquark pair).
\end{enumerate}

This algorithm gives the following expression for
the probability for decay of the given cluster (\texttt{ID1,ID3})
into the hadrons with codes \texttt{IR1=NCLDK(I1)} and 
\texttt{IR2=NCLDK(I2)}
\begin{equation}
  \label{P1}
  P(\texttt{IR1},\texttt{IR2},\texttt{ID2}|\texttt{ID1},\texttt{ID3})
  = 
  P_{flavour}\cdot P_{spin}(\texttt{IR1})
  \cdot P_{spin}(\texttt{IR2})
  \cdot P_{phase}\ . 
\end{equation}

It is now clear, why the inclusion of some new massive resonances with
a given flavour leads to the suppression of all decay channels with
the hadrons of the same flavour. Omitting hadron spins and mixing
weights, probability (\ref{Ps}) depends only on the number of
iso-flavour particles
\begin{equation}
  \label{Ps2}
  P_{spin}(\texttt{IR1}|\texttt{ID1}, \texttt{ID2})=
  \frac{1}{\texttt{RESN(ID1,ID2)}}\ .
\end{equation}
Finally, the overall probability of a cluster to decay into channels with created
flavour \texttt{ID2} is
\begin{eqnarray}
  \nonumber
  P(\texttt{ID2}|\texttt{ID1},\texttt{ID3}) &=& 
  \sum_{\texttt{IR1},\texttt{IR2}} 
    P(\texttt{IR1},\texttt{IR2},\texttt{ID2}|\texttt{ID1},\texttt{ID3})\\
 \label{P2}
 &\propto&
  \texttt{PWT(ID2)}\cdot\frac{\sum_{\tiny{\texttt{IR1},\texttt{IR2}}}P_{phase}}
  {\texttt{RESN(ID1,ID2)}\cdot\texttt{RESN(ID2,ID3)}}\ ,
\end{eqnarray}
which is nothing else than the \textbf{mean} value of phase spaces of the
corresponding decay channels. The phase space of
resonances is small due to their high masses. Hence, this new
resonances lower value of the overall probability~(\ref{P2}).

The motivation for the {\sc Herwig} authors was to eliminate factor
$1/2$ in $u\bar{u}$ and $d\bar{d}$ mixing for neutral non-strange
mesons,  \cite{Bryan}. Indeed, this algorithm gives the same number of
$\pi^0$ and $\pi^+$ in the case where there are only $u$ and $d$
quarks, with no baryons, and with $\pi^0$, $\pi^\pm$, and
isotopic ($\eta$-like) singlet.

This behaviour of \HER\  is demonstrated in Tab.~\ref{Tab1}. Both
versions (5.8 and 5.9) have the same default settings but \HER\
particle table contains more meson resonances. Consequently, baryon
production was enhanced and meson production decreased.
\section{New cluster hadronization model}

In the rest frame of a particle of mass $M$, the width of a two-body decay is given, generally, by expression
\begin{equation}
  \label{Decay}
  \dd\Gamma=\frac{1}{32\pi^2}\left|\mathcal{M}\right|^2
            \frac{\left|\mbox{\boldmath $p_1$}\right|}{M^2}\dd\Omega\,,
\end{equation}
where final particles are labeled 1 and 2, $\mathcal{M}$ is the corresponding matrix element, and 
\begin{equation}
  \label{PCM}
  \left|\mbox{\boldmath $p_1$}\right|=\left|\mbox{\boldmath $p_2$}\right|=
  \frac{\left[\left(M^2-\left(m_1+m_2\right)^2\right)
              \left(M^2-\left(m_1-m_2\right)^2\right)\right]^{1/2}}{2M}=
  \frac{\texttt{PCM}\left(M,m_1,m_2\right)}{2M}
\end{equation}
are the momenta of final state particles.

We do not know a great deal about the matrix element. We can sum over final
states (spin factors), we can take into account mixing of flavours
(mixing weights) and we can apply Zweig's rule which states the two-body
decays of cluster \texttt{(ID1,ID3)} to pair of hadrons with flavour
\texttt{(ID1,ID3)} and \texttt{(ID2,ID2)} are strongly
suppressed. Assuming that the matrix elements are the same for all decay 
channels of a cluster and that they depend only on the flavour \texttt{ID2} of
a created (di)quark anti(di)quark pair, the weight
$W(\texttt{IR1,IR2,ID2|ID1,ID3})$ of particular decay of cluster
\texttt{(ID1,ID3)} into hadrons \texttt{IR1} and \texttt{IR2} with
flavour content \texttt{(ID1,ID2)} and \texttt{(ID2,ID3)} is
\begin{eqnarray}
  &&\hspace*{-0.5cm}W(\texttt{IR1},\texttt{IR2},\texttt{ID2}|\texttt{ID1},\texttt{ID3})=
    \texttt{PWT(ID2)}\cdot(2J_1+1)\cdot(2J_2+1)\cdot
     \texttt{WT(IR1)}\cdot\texttt{WT(IR2)}\cdot\nonumber\\
  &&\label{W1}\hspace*{4.57cm}\cdot\,
    \texttt{SWTEF(IR1)}\cdot\texttt{SWTEF(IR2)}\cdot
    \texttt{PCM(EM0,EM1,EM2)}\ ,
\end{eqnarray}
where $J_1$, $J_2$ are particle spins and \texttt{WT} is flavour
mixing weight.

The overall probability for decay channels with created flavour
\texttt{ID2} is then (again omitting spins, mixing weights, and
individual weights \texttt{SWTEF}) 
\begin{equation}
  \label{W2}
  P(\texttt{ID2}|\texttt{ID1},\texttt{ID3})\propto
  \texttt{PWT(ID2)}\cdot\!\sum_{\texttt{IR1,IR2}}\texttt{PCM}\ .
\end{equation}
Contrary to Eq.(\ref{P2}) this probability is proportional to the
\textbf{sum} of the phase spaces of particular channels. Therefore, the
inclusion of new baryon resonances will enhance the baryon
multiplicities.

The new hadronization model is based on formula~(\ref{W1}). Several
subroutines have been changed to implement this model into \HER\
(\texttt{HWCHAD, HWURES}). Moreover, \HER\  particle table had to be
extended. As was mentioned earlier, there are plenty of meson
resonances in the table but only the basic baryon octet and decuplet. New
baryon resonances dramatically increase the probability for a
cluster decay through the baryon channel and they play an important
role in the new model. Hence, I decided to include all light quark
meson and baryon resonances that have their own PDG number
in~\cite{PDG94}. The branching ratios of particular decay modes were
taken also from this paper. The properties of these new resonances
were put in new subroutine \texttt{MHWRES}.
\subsection{Tuning \HER\  with new hadronization model}

To make final comparison between the both hadronization models, one
has to use the tuned settings of \HER\  parameters. The L3 fit has
been used for the original model (see Tab.~\ref{setting}).

%
%
\begin{table}[htbp]
  \begin{center}
    \leavevmode
    \begin{tabular}{|l|c|c|c|}
      \hline & H5.9 & H5.9n & H5.9n\\
      \hline parameter &\bfseries{L3 fit}&\bfseries{CAND1}&\bfseries{CAND2}\\
      \hline \texttt{QCDLAM}    & 0.177 & 0.179 & 0.1602 \\
      \hline \texttt{RMASS(13)} & 0.75  & 0.706 & 0.764  \\
      \hline \texttt{CLMAX}     & 3.006 & 5.27  & 3.90   \\
      \hline \texttt{CLPOW}     & 2.033 & ---   & 1.53   \\
      \hline \texttt{CLSMR}     & 0.35  & ---   & 0.341  \\
      \hline \texttt{DECWT}     & 0.5   & ---   & 0.753  \\
      \hline \texttt{PWT(3)}    & 0.88  & ---   & 1.071  \\
      \hline \texttt{PWT(7)}    & 0.80  & 0.812 & 2.408  \\
      \hline
    \end{tabular}
    \caption{Settings of \HER\  for both hadronization models.}
    \label{setting}
  \end{center}
\end{table}

Complete tuning of \HER\  with new hadronization model has not been
attempted.  I used the same method as in~\cite{DelphiFit} but with
fewer statistics so the presented results can still be improved.
DELPHI measurements of event shape variables~\cite{DelphiFit} and LEPI
data of $p$--spectra of identified particles have been used for tuning
the parameters. There are two settings for \HER\  with new
hadronization (H5.9n) in the Tab.~\ref{setting}. In the case of {\bf
CAND1} setting, only four parameters have been tuned. {\bf CAND2}
represents the best fit where the same parameters was tuned as in the
case of L3 fit. Tab.~\ref{FitRes} gives a summary of the fit results.
%
%
\begin{table}[htbp]
  \begin{center}
    \leavevmode
    \begin{tabular}{|l|l|l|l|}
      \hline & Event shapes & Ident. particles & All channels\\
      \hline \bfseries{L3 fit} & $\chi^2=2874/305=9.4$ 
                               & $\chi^2=1002/157=6.4$
                               & $\chi^2=3876/462=8.4$\\
      \hline \bfseries{CAND1}  & $\chi^2=3121/260=12.0$
                               & $\chi^2=921/130=7.1$
                               & $\chi^2=4042/390=10.5$\\
      \hline \bfseries{CAND2}  & $\chi^2=2319/305=7.6$
                               & $\chi^2=824/157=5.2$
                               & $\chi^2=3143/462=6.8$\\
      \hline
    \end{tabular}
    \caption{Summary of fit results.}
    \label{FitRes}
  \end{center}
\end{table}
\section{Comparison of the two cluster hadronization models}

Although the probability of cluster decay through the particular
channel is proportional to the phase space in both cases, the strange
normalisation factors of the standard version (see Eq.~(\ref{Ps}))
give different behaviour of the models.

{\bf CAND1} setting of \HER\  is the worst one (see
Tab.~\ref{FitRes}). This could be due to a small number of
fitted parameters. Nevertheless, one can find quite a good agreement
with the identified particle multiplicities data in the basic meson
and baryon octet, Tab.~\ref{Tab1}.
%
%
\begin{table}[p]
\renewcommand{\arraystretch}{1.1}
\begin{center}
\begin{tabular}{|l|%
                r@{$\,\pm\,$}l|%
                c|c|c|c|c|c|
               }
  \hline
    \bfseries Particle &
    \multicolumn{2}{|c|}{\bfseries PDG'98} &
    \sc H5.8&
    {\sc H5.8}&
    {\sc H5.9}&
    {\sc H5.9}&
    {\sc H5.9}n&
    {\sc H5.9}n
  \\
  \hline  \multicolumn{3}{|r|}{}&\bf Delphi& \bf DEF&
               \bf DEF&\bf L3& \bf CAND  & \bf CAND\\
  \hline  \multicolumn{3}{|r|}{}&\bf FIT& 
               & &  \bf FIT& \bf ONE & \bf TWO\\
  \hline {}&\multicolumn{2}{|c|}{}&{}&{}&{}&{}&{}&{}\\[-4.9mm]
  \hline charged& 21.05&$0.20^{\left[8\right]}$& 20.81 &21.17& 20.62&
               21.13& 21.03 & 21.52 \\
  \hline {}&\multicolumn{2}{|c|}{}&{}&{}&{}&{}&{}&{} \\[-4.9mm]
  \hline $\pi^0$& 9.42&0.56& 9.79& 9.76& 10.14& 10.88& 9.51 & 9.74 \\
  \hline $\pi^\pm$& 17.1&0.4& 17.64& 17.52& 16.7& 17.62& 17.26& 17.53\\
  \hline $K^0$& 2.013 & 0.033 & 2.03& 2.29& 1.937& 1.949&1.981&2.253\\
  \hline $K^\pm$ & 2.39 & 0.12 & 2.11& 2.40& 2.10& 2.20&2.32&2.54\\
  \hline $\eta$ & 0.97 & 0.10 & 1.01& 1.01& 0.92& 1.01&0.79&0.89\\
  \hline $\eta^\prime$ & 0.222 &0.040&0.14&0.144&0.140&0.157&0.109&0.125\\
  \hline {}&\multicolumn{2}{|c|}{}&{}&{}&{}&{}&{}&{} \\[-4.9mm]
  \hline $\rho(770)^0$ & 1.28 & 0.14 & 1.43& 1.32& 1.16& 1.29&1.12&1.12\\
  \hline $K^*(892)^0$ & 0.747 & 0.028 & 0.75& 0.796& 0.521&0.565&0.640&0.734\\
  \hline $\Phi(1020)$ & 0.109 & 0.007 & 0.099& 0.120& 0.181&0.178&0.208&0.182\\
  \hline $\omega(782)$ & 1.10 & 0.13 & 0.92& 1.10& 1.05 & 1.23& 1.00& 1.02\\
  \hline {}&\multicolumn{2}{|c|}{}&{}&{}&{}&{}&{}&{}\\[-4.9mm]
  \hline $p$ & 0.964 & 0.102 & 0.77& 0.948& 1.41 & 0.77&0.959&0.944 \\
  \hline $\Lambda$ & 0.372 & 0.009 & 0.367& 0.431& 0.598 &0.256&0.363&0.389 \\
  \hline $\Sigma^-$ & 0.071 & 0.018 & 0.058& 0.066& 0.10 & 0.069&0.082&0.092 \\
  \hline $\Xi^-$ & 0.0258 & 0.0010 & 0.048& 0.052&0.070&0.0242&0.0195&0.0208\\
  \hline {}&\multicolumn{2}{|c|}{}&{}&{}&{}&{}&{}&{}\\[-4.9mm]
  \hline $\Delta^{++}$ & 0.085 & 0.014 & 0.158& 0.198&0.276&0.109&0.113&0.100\\
  \hline $\Sigma(1385)^\pm$ & 0.0462 & 0.0028 & 0.118& 0.15 & 0.20 &
               0.069& 0.0713&0.0617 \\
  \hline $\Xi(1530)^0$ & 0.0055 & 0.0005 & 0.026& 0.024 & 0.037 &
               0.0053& 0.0120&0.0091 \\
  \hline $\Omega^-$ & 0.0016&0.0003&0.0074&0.0078&0.0094&0.0009&0.0019&0.0011\\
  \hline {}&\multicolumn{2}{|c|}{}&{}&{}&{}&{}&{}&{} \\[-4.9mm]
  \hline $\Lambda(1520)$ & 0.0213 & $0.0028^{\left[8\right]}$& --- &
               --- & --- & --- & 0.0426&0.0339\\
  \hline
\end{tabular}
\end{center}
\caption{Particle multiplicities per event in \EEBAR\
  interactions at $\sqrt{s}=91.2\,\mathrm{GeV}$ in the
  data~\cite{PDG98} compared with {\sc Herwig~5.8} (H5.8), \HER\
  prediction for default cluster  hadronization model (H5.9),  and for
  new hadronization model (H5.9n).}
\label{Tab1}
\end{table}

On the other hand, {\bf CAND2} gives a better description of the data
than the \HER\  default model with the {\bf L3} setting. The set of
fitted parameters was the same, so it is more appropriate to compare
these two settings. Since the {\bf L3} fit was made on slightly
different data some inconsistency still survives.

Concerning identified particle multiplicities, {\bf CAND1} predictions
are in an excellent agreement with the data in the whole meson sector
except $\eta^\prime$ and $\Phi$. Good agreement is also found in the
basic baryon octet (except $\Xi^-$) but large discrepancies occur in
the basic baryon decuplet ($\Sigma(1385)$ and $\Xi(1530)^0$).

The situation is roughly the same in the case of {\bf CAND2} set. The
agreement with the data is slightly worst (especially for $K^0$ and
$K^\pm$) and the problems with baryon resonances were not cured
although parameter \texttt{DECWT} was also tuned in this fit.

The default \HER\  hadronization model ({\bf L3} fit) has similar
problems in describing the data here. Moreover, it underestimates
the production of protons and $\Lambda$ particles.

Both hadronization models give (within the experimental uncertainties)
the correct particle multiplicities for the lightest non-strange
mesons and baryons ($\pi$, $\eta$, $\rho$, $\omega$, $p$, $\Delta$)
although the new model predictions are slightly closer to the
data. The discrepancies occur for the singlet mesons $\eta^\prime$ and
$\Phi(1020)$ and for heavy strange baryons.
\section{Conclusions}

It has been observed that with the introduction of new nucleon
resonances, \HER\  predicts a reduced proton yield, contrary to the
naive view of cluster hadronization. In the \HER\  hadronization
scheme, a decay of a colourless cluster takes place in such a way as
to eliminate the factor $1/2$ in $u\bar{u}$ and $d\bar{d}$ mixing. The
algorithm, however, gives the probability of a cluster decay through
channels with a created particular flavour that is proportional to the mean
value of corresponding phase spaces.

A new hadronization model which treats all decay channels in the same
way has been proposed. It was found that this provides an improved
description of inclusive particle data taken by the LEP
experiments. In particular, the baryon sector is better described that
with the default version of \HER. This is a particularly important if
the {\sc Herwig} model is to be used predictively in HERA environment,
in which topics such as baryon number propagation and strange diquark
suppression are starting to be studied.

\end{document}